\documentclass[11pt, letterpaper]{article}

\usepackage{graphicx}
\usepackage{amsmath}
\usepackage{amssymb}
\usepackage{setspace}
\usepackage{epstopdf}
\usepackage{color}
\usepackage[letterpaper,left=1in,right=1in,top=1in,bottom=1in]{geometry}
\usepackage[linesnumbered,ruled,vlined,longend]{algorithm2e}
\usepackage{multirow}
\usepackage{longtable}
\usepackage{rotating}
\usepackage{threeparttable}
\usepackage{colortbl}
\usepackage{amsthm}
\usepackage[modulo]{lineno}
\usepackage{enumerate}

\newtheorem{remark}{Remark}

\newtheorem{proposition}{Proposition}

\newtheorem{definition}{Definition}

\newcommand{\m}{\mathbb}

\title{\Large \bf An efficient implementation of graph-based invariant set algorithm for constrained nonlinear dynamical systems}

\author{\centerline{\normalsize Benjamin Decardi-Nelson$^{a}$, Jinfeng Liu$^{a,}$\thanks{Corresponding author: J. Liu. Tel: +1-780-492-1317. Fax: +1-780-492-2881. Email: jinfeng@ualberta.ca}}\vspace{5mm}\\
    \centerline{\small $^{a}$ Department of Chemical \& Materials Engineering, University of Alberta,}\\
    \centerline{\small Edmonton, AB, Canada, T6G 1H9}}

\allowdisplaybreaks

\begin{document}

\date{}

\maketitle
\setstretch{1.39}

\begin{abstract} 

The graph-based invariant set (GIS) algorithm is a promising set-based technique for computing the largest (with respect to inclusion) control invariant set of general discrete-time nonlinear dynamical systems. However, like other invariant set algorithms for nonlinear systems, the GIS algorithm may require a lot of resources when computing the control invariant set. This limits its applicability to higher dimensional systems. In this work, we present an improved and efficient implementation of the GIS algorithm for general discrete-time controlled nonlinear systems. We first identify the bottlenecks through extensive analysis, and then provide remedial procedures to improve the implementation of the GIS algorithm. Specifically, we developed an adaptive subdivision scheme using a supervised machine learning-based algorithm to reduce the cell growth rate and parallelize the graph construction step. We extensively demonstrate the performance of the improved GIS algorithm using a numerical example and compare the result to that of the standard GIS algorithm. The results show that the adaptive subdivision and the parallelization improved the speed of the algorithm by about 8x and 3x respectively, that of the standard GIS algorithm.

\end{abstract}


\section{Introduction}\label{sec:introduction}

Set invariance theory plays a fundamental role in constrained control systems design and analysis. It has been found to be particularly useful in the design of model predictive control algorithms to ensure constraint satisfaction, recursive feasibility and stability \cite{mayne2001,cannon2003}. For these reasons, control invariant sets (CISs) have received significant attention in the systems and control literature \cite{blanchini1999}. A set is said to be control invariant if trajectories that start from it, can be forced to remain in it by using admissible control inputs. It is worth mentioning that control invariant sets are closely linked to viability theory \cite{aubin2009,maidens2013}, reachability analysis \cite{mitchell2005,lygeros2004} and null controllability \cite{homer2017,homer2018,homer2020}. Unfortunately, determining control invariant sets is a very challenging task, even for linear systems. 

To this end, substantial efforts have been devoted over the past decades to determine control invariant sets. Linear systems have, in particular, received significant attention in literature. Several results which address computational issues and algorithmic procedures exist for both deterministic and uncertain linear systems \cite{rungger2017,rakovic2005,kerrigan2001,kolmanovsky1998,gilbert1991}. For general nonlinear systems however, only a few results exit \cite{fiacchini2010,alamo2009,bravo2005,homer2018,homer2020}. Although Lyapunov functions are important tools for set invariance analysis, it is difficult to obtain such functions for nonlinear systems in general. Moreover, while most algorithms for linear systems provide convergence to the largest CIS, this is not the case for nonlinear systems. Recent advances in safe reinforcement learning (RL) \cite{berkenkamp2017,chow2018,FISAC2019}  and zone model predictive control (ZMPC) \cite{DECARDI_NELSON2022,liu2019,liu2018} further motivate the need to determine the largest CIS.


Recently, the graph-based invariant set (GIS) algorithm has been successfully used to determine the control invariant set of complex nonlinear dynamical systems \cite{decardi2021}. More importantly, convergence to the largest CIS was also provided. In the GIS algorithm, the dynamics of the system is approximated with a directed graph and then analyzed to obtain an approximation of the largest CIS. However, the GIS algorithm, like other control invariant set algorithms for nonlinear systems, may require high computing resources. This limits the applicability of the algorithm to high dimensional systems.

In this article, we present details of an improved and efficient implementation of the GIS algorithm for computing control invariant sets of constrained dynamical systems. Obviously, only the boundary of the CIS is of interest during the computation. This is the central idea we employ in the improved GIS algorithm. Our approach involves an adaptive subdivision technique to slow down the cell growth rate, and parallelization of the graph construction with multicore processing and graphics processing units (GPU). The adaptive subdivision technique makes use of a supervised machine learning technique to select the cells for subdivision. We compare the results obtained by the improved GIS algorithm to that of the standard GIS algorithm using a nonlinear example. 

The rest of the paper is organized as follows: In Section \ref{sec:gis_algorithm}, we briefly present the standard GIS algorithm with a focus on the computational requirements. Section \ref{sec:efficient_gis_algorithm} explores the efficient implementation of the GIS algorithm by providing effective remedial actions for each part of the algorithm. Section \ref{sec:results} presents the results of the improved algorithm using a numerical example and we provide some concluding remarks in Section \ref{sec:concluding_remarks}.

\noindent \textbf{Notation.} $\mathbb{Z}$ denotes the set of integers $\{\hdots,-2,-1,0,1,2,\hdots \}$. $\mathbb{Z}_+$ denotes the set of non-negative integers $\{ 0,1,2,\hdots \}$. $\{ z_k \}_{k \in \mathbb{Z}_+}$ denotes an ordered set of numbers according to $k \in \mathbb{Z}_+$ $\{ z_0, z_1, z_2, \hdots \}$. 
The operator $|\cdot|$ denotes the Euclidean norm of a vector. 
A directed graph is denoted as $G=(V,E)$ with $V$ denoting the set of vertices of the graph and $E$ denoting the set of ordered pairs of vertices known as edges. 

\section{The standard GIS algorithm}\label{sec:gis_algorithm}

In this section, we briefly present the graph-based invariant set (GIS) algorithm for computing the largest (with respect to inclusion) control invariant set $R_X$ of general controlled nonlinear dynamical systems. The reader may refer to study of Decardi-Nelson and Liu \cite{decardi2021} for a more detailed discussion on this subject.

We consider discrete-time nonlinear systems of the form
\begin{equation}\label{eqn:system}
  x^+ = f(x,u)
\end{equation}
where $x^+ \in X \subseteq \mathbb{R}^{n}$ denotes the state at the next sampling time, $x \in X \subseteq \mathbb{R}^{n}$ is the state, $u \in U \subseteq \mathbb{R}^{m}$ represents the control input. The sets $X$ and $U$ denote the state and input constraints respectively. We assume that the sets $X$ and $U$ are compact, and the function $f:X \times U \rightarrow X$ is a sufficiently smooth vector field in $X$.

We begin this section by describing the three major steps in the standard GIS algorithm, namely cell subdivision, graph construction and graph analysis. Thereafter, we briefly analyze the computational requirements for each step of the algorithm. 


\subsection{Cell subdivision}

\begin{figure}[tbp] 
    \center{\includegraphics[width=0.4\columnwidth]{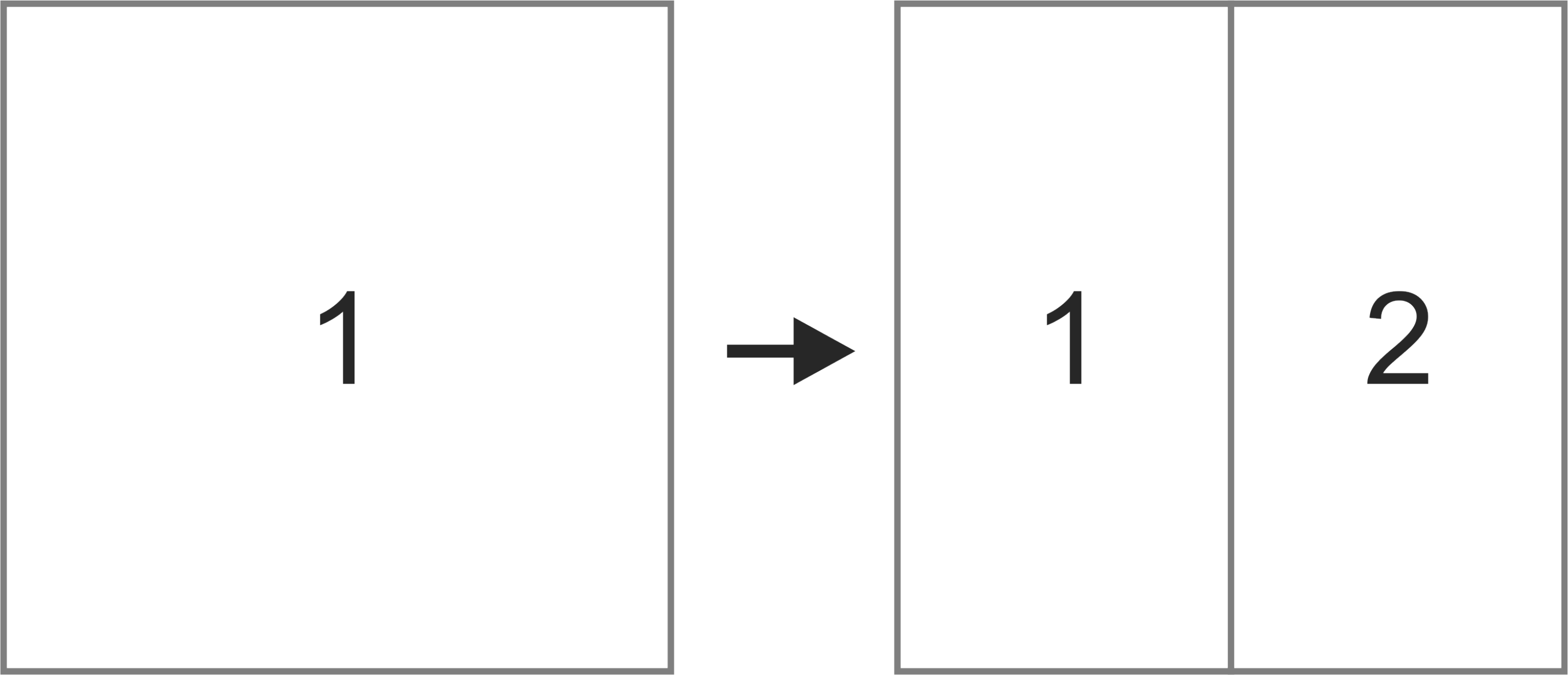}}
    \caption{\label{fig:subdivision} Cell subdivision process at each iteration of the algorithm }
\end{figure}

For improved computational efficiency, the algorithm makes use of the subdivision technique \cite{dellnitz2002} during the approximate computation of $R_X$. This leads to the generation of the sequence $\mathcal{C}_{d_0}, \mathcal{C}_{d_1}, \mathcal{C}_{d_2}, \cdots$ of finite collections of closed sets known as cells $B_i,i=1,2,\cdots,l$ with the property that for all positive integers $k$, $R_k = \cup_{B_i \in \mathcal{C}_{d_k}}$ is a covering of the largest control invariant set $R_X$. The sequence of coverings is constructed in such a way that the diameter of the covering $d_k$
\begin{equation*}
    d_k = \text{diam}(\mathcal{C}_{d_k}) := \max_{B_i \in \mathcal{C}_{d_k}}\text{diam}(B_i)
\end{equation*}
where diam($B_i$) = $\text{sup}\{ |x-y|:x,y \in B_i \}$, converges to zero as $k \rightarrow \infty$. In the subdivision step, a finer covering of the RCIS is generated by dividing the current cells along one of the dimensions. If $\mathcal{C}_{d_k}$ and $\mathcal{C}_{d_{k-1}}$ are coverings of the $R_X$ where $d_k$ and $d_{k-1}$ denote their respective diameters, then $d_k < d_{k-1}$ and    
    \begin{equation*}
        \cup_{B \in \mathcal{C}_{d_k}} B = \cup_{B \in \mathcal{C}_{d_{k-1}}} B
    \end{equation*}
The set that is subdivided does not change other than having cells with smaller diameter. In each iteration of the algorithm, the dimension along which the cells are divided is cycled. An illustration of the cell subdivision is shown in Figure~\ref{fig:subdivision}.

\subsection{Graph construction}

Given System~\eqref{eqn:system} with the associated constraints on the state and input, the largest control invariant set $R_X$ can be computed by first approximating the dynamics of the system using a directed graph.  Thereafter, an approximation of the control invariant set may be obtained by analyzing the resulting directed graph. 

Let us rewrite System~\eqref{eqn:system} in a form which makes it easier for the graph construction. To achieve this, System \eqref{eqn:system} is parameterized with the input $U$ to obtain a set-valued map of the form
\begin{equation} \label{eqn:differential_inclusion}
    F(x) := f(x,U) =  \{f(x,u) \}_{\cup_{u \in U}}
\end{equation}
The map $F$ associates with each state $x$ to the subset $F(x)$ of feasible next states. Therefore, System~\eqref{eqn:system} defined by the family of parameterized difference equations is actually governed by the difference inclusion

\begin{equation}\label{eqn:difference_inclusion}
    x^+ \in F(x)
\end{equation}

\begin{definition}[Symbolic image \cite{osipenko2007}]
    Let $G$ be a directed graph with $l$ vertices where each vertex is a cell or box $B_i$ in a finite covering $\mathcal{C}$ of the domain $X$ of System~\eqref{eqn:difference_inclusion}. The vertices $B_i$ and $B_j$ are connected by a directed edge $B_i \rightarrow B_j$ if 
    \begin{equation*}
        B_j \cap F(B_i) \neq \emptyset
    \end{equation*}
    where $F(B_i):=\{y | y= F(x), x \in B_i\}$. The graph $G$ is called a symbolic image of System~(\ref{eqn:system}) with respect to the covering $\mathcal{C}$.
\end{definition}

\begin{definition}[Admissible path \cite{osipenko2007}] \label{admissible_path}
    A sequence $\{ z_k \}_{k\in \mathbb{Z}_+}$ with each element $z_k$ taking a value from the set of vertices of $G$ is called an admissible path if for each $k \in \mathbb{Z}_+$, the graph $G$ contains the edge $z_k \rightarrow z_{k+1}$.
\end{definition}

\begin{definition}[Out-degree of a vertex]
        The out-degree of a vertex in a directed graph is the number of edges going out of the vertex.
\end{definition}

An admissible path on the symbolic image $G$ may either be finite or infinite. An admissible path on the symbolic image is finite if it ends with a vertex that has zero out-degree. Otherwise, it is infinite. To understand the relationship between an admissible path on the symbolic image and the trajectories of System~(\ref{eqn:system}), we recall the notion of $\varepsilon$-orbit.

\begin{definition}[$\varepsilon$-orbit \cite{sakai1994}]
    For a given $\varepsilon > 0$, a sequence of points $\{ x_k \}_{k \in \mathbb{Z}_+}$ in $X$ is called an $\varepsilon$-orbit of System (\ref{eqn:system}) if for any $k \in \mathbb{Z}_+$
    \begin{equation*}
        |f(x_k, u_k) - x_{k+1}| < \varepsilon 
    \end{equation*}
\end{definition}

Trajectories of a dynamical system computed by a computer are $\varepsilon$-orbit for a sufficiently small $\varepsilon$ as a result of round off errors. Real trajectories are seldom known in practice. There is a natural correspondence between admissible paths on the symbolic image and the $\varepsilon$-orbits. An admissible path on the symbolic image represents a $\varepsilon$-orbit and vice versa. This means that if the sequence $\{ z_k \}_{k \in \mathbb{Z}_+}$ is an admissible path on the symbolic image $G$, then there exist the sequence $\{ (x_k, u_k), x_k \in z_k, u_k \in \m{U} \}_{k \in \mathbb{Z}_+}$ that is an $\varepsilon$-orbit of system (\ref{eqn:system}) such that the following inequality holds

 \begin{equation*}
    |f(x_k, u_k)-x_{k+1}| \leq \text{diam}(z_{k+1}) < \varepsilon
 \end{equation*}
It is easy to see that the finer the covering, the more precise the approximation of the system trajectories on the symbolic image. 





\subsection{Analysis on the directed graph}

In this subsection, we describe how the resulting directed graph can be investigated using graph theory to obtain an outer approximation of the largest control invariant set for dynamical systems. 

Let $R_k$ obtained from the collection $\mathcal{C}_{d_k}$ be a covering of $R_X$. If a vertex ($B_i$) of the symbolic image of system (\ref{eqn:system}) has zero out-degree, then its image $F(B_i)$ has no intersection with any other vertex on the symbolic image; i.e., $F(B_i) \cap R_k = \emptyset$. Therefore its image $F(B_i)$ lies outside the covering of $R_X$. This implies that any trajectory starting from the cell will exit $R_k$ and ultimately exit the state constraint $X$ in finite time. Specifically, it implies that there does not exist an input from the admissible input set $U$ that is able to keep the state in $X$. 

 \begin{definition}[Strongly connected graph] \label{strongly_connected}
     A directed graph $G=(V,E)$ is said to be strongly connected if there is an admissible path in both directions between each pair of vertices of the graph.
 \end{definition}

For a graph $G$ that is not strongly connected, it may contain subgraphs that are strongly connected. These subgraphs are known as the strongly connected component subgraphs of $G$. 

\begin{definition}[Non-leaving cells] \label{defn:nonleaving}
    The set of vertices of the directed graph $G$ with infinite admissible paths passing through them, denoted as $I^+(G)$, is the union of vertices of the largest strongly connected component subgraph of the directed graph $G$ and any vertex of $G$ that is not in the largest strongly connected components but has a path to a vertex in the largest strongly connected component subgraph.
\end{definition}

The overall goal after constructing the graph is therefore to use graph tools to identify the non-leaving cells on the symbolic image. The following proposition summarizes how we may obtain an outer approximation of the largest control invariant set of system (\ref{eqn:system}) based on its symbolic image \cite{decardi2021}.

 \begin{proposition} \label{prop:nonleaving_control}
     Consider System \eqref{eqn:system} with its associated state $X$ and input $U$ constraints. Let $G=(V,E)$ having a set of vertices $V$ and a set of ordered pairs of vertices $E$ be a symbolic image of the difference inclusion $F$ in (\ref{eqn:difference_inclusion}) with respect to a finite covering $\mathcal{C}$ of $X$. Then 
     \begin{enumerate}[i.]
        \item \label{theorem_1_i} the vertices of the largest strongly connected component subgraph $G_s=(V_s,E_s)$ of $G$ have infinite admissible paths passing through them.
        \item \label{theorem_1_ii} any element of $V$ but not $V_s$ with a path to at least one vertex of $G_s$ also has an infinite admissible path passing through it.
        \item \label{theorem_1_iii} the union of the elements of $(i)$ and $(ii)$, $I^+(G)$, is a closed neighbourhood of the largest forward invariant set $R_X$ of System (\ref{eqn:system}) in $X$; that is,
     \begin{equation}
        R_X\subseteq I^+(G)
     \end{equation}
     \end{enumerate}
 \end{proposition}

\subsection{Computational requirements of standard GIS algorithm}








When the standard GIS algorithm is used to compute the largest CIS, the computational requirements are directly proportional to the number of cells used to approximate the set. This implies that as the number of cells increase at each iteration, the algorithm gets slower while the memory needed goes higher. Thus, the algorithm is greatly influenced by the number of cells used to approximate $R_X$. Let $n_c$ be the number of cells at iteration $k$ of the algorithm. The subdivision step involves iterating through all the cells to divide each cell into two. This results in a linear time complexity, that is $O(n_c)$. The graph construction step involves (1) finding one step forward mappings of each cell, (2) finding which cells intersect the one step forward mapping and (3) constructing an edge list of the graph. Without going into the details of how the one step mapping is achieved for each cell, the time complexity is $O(n_c)$ since the one step forward mapping need to be created for each cell. Finding the cells that have an intersection section with the one step forward mapping involves using an R*-tree. The average case (because we do not have data overlaps) time complexity of querying the R*-tree for each cell is $O(\text{log }n_c)$. Hence, the time complexity for finding the intersection of the one step forward mapping of all the cells is $O(n_c \text{log }n_c)$. Finally, the complexity of creating the edge list of the digraph is $O(|E|)$ where $|E|$ is the number of edges on the digraph. The overall time complexity of the graph construction step is $O(n_c \text{log }n_c + |E|)$. Analyzing the digraph involves finding the non-leaving cells. This has a time complexity of $O(|V|+|E|)$ where $|V|$ is the number of vertices which is equivalent to $n_c$. The time complexities of the major parts of the GIS algorithm as well as their significance are presented in Table \ref{tb:gis_complexity_time}.


\begin{table}[tp!]
  \begin{center}
  \caption{Major parts of the GIS algorithm and their computational requirements }\label{tb:gis_complexity_time}
  \vspace{2mm}
    \begin{tabular}{lll}
        \hline
    Part & Time complexity & Significance \\  
        \hline
    Cell subdivision & $O(n_c)$ & Not significant \\ 
    Graph construction  & $O(n_c \text{log }n_c + |E|)$ & Significant \\ 
    Graph analysis & $O(|V|+|E|)$ & Not significant \\
        \hline
    \end{tabular}
  \end{center}
\end{table}

\section{The improved and efficient GIS algorithm}\label{sec:efficient_gis_algorithm}

As mentioned earlier, the standard GIS algorithm approximates the largest control invariant set by iteratively refining the cells that cover $R_X$. In doing so, it may over refine cells which may be dynamically irrelevant to the computation. This is because in the subdivision step of the standard algorithm, each cell is divided. While this works in principle, it may generate high number of cells as the algorithm proceeds. This significantly slows down the algorithm, and lead to high memory storage and computational requirements as demonstrated in the preceding section. In this section, we propose a method to improve the computational efficiency of the algorithm by adaptively selecting a subset of the cells to be subdivided instead of subdividing all the cells. This reduces the cell growth rate and ultimately the overall computational requirements of the algorithm. In addition, we describe a parallel implementation of the graph construction step to speed up the standard algorithm using graphics processing unit (GPU).

We begin this section by first describing the adaptive subdivision method. Thereafter, we describe the parallelization of the graph construction step using GPU. Finally, we briefly discuss the implications of the modifications on the convergence of the GIS algorithm to $R_X$.

\subsection{Adaptive cell subdivision}

\begin{figure}[tbp] 
    \center{\includegraphics[width=0.7\columnwidth]{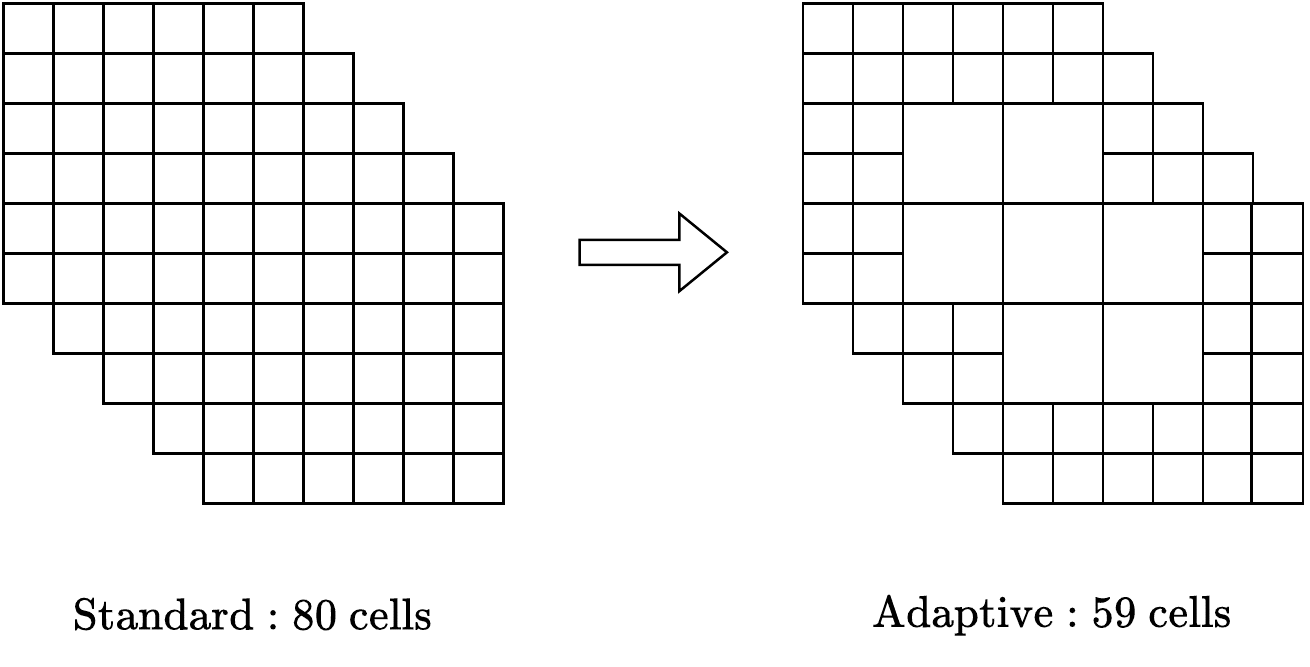}}
    \caption{\label{fig:adaptive_cell_division} Representation of the same set with different number of cells. Left: Uniform cell subdivision as used in the standard algorithm. Right: Adaptive subdivision where the boundary is refined and the interior is not}
\end{figure}

Theoretically, the part of the control invariant set of interest to us is its boundary. For linear systems with convex state and input constraints, the largest control invariant set is convex. This explains why for linear systems with convex constraints, it is sufficient to test for control invariance only at the vertices (boundary) as the iteration progresses. For nonlinear systems however, the largest control invariant set may not be convex. The set can be in any form or shape depending on the system dynamics. Moreover, there could be more than one invariant set in the region of interest. Hence, the search for the control invariant set involves searching everywhere within the state constraint. 
\begin{figure}[tbp] 
    \center{\includegraphics[width=0.5\columnwidth]{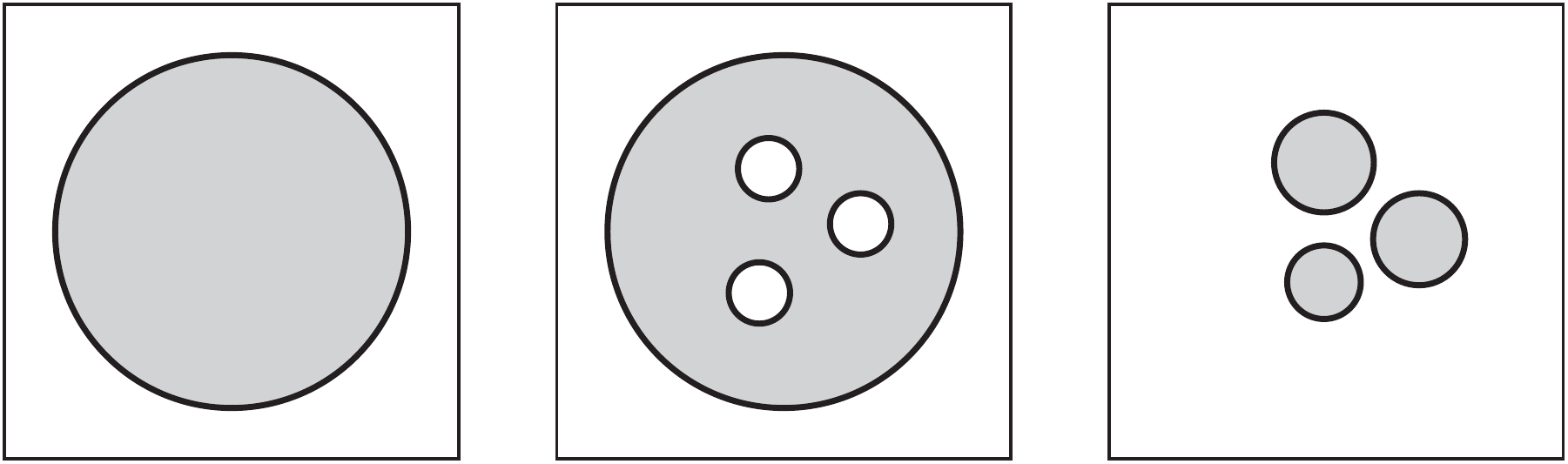}}
    \caption{\label{fig:adaptive_cis_boundary_types} Types of control invariant set boundaries. The thick black lines represent the boundary of the set and the shaded portion represent the interior of the control invariant set. The box represent the region of the state space of interest $X$. Left:  Control invariant set with continuous boundary; Middle: Control invariant set with pocket of holes creating a discontinuous boundary; Right: Multiple control invariant set in the search region.}
\end{figure}
Figure \ref{fig:adaptive_cis_boundary_types} shows some examples of the shape of control invariant set that may be encountered in nonlinear systems. If the boundary area of the control invariant set contained in $X$ is roughly known, then it suffices to just refine around the boundary area and use cells with bigger diameter for the interior. As an example, Figure \ref{fig:adaptive_cell_division} shows how the same set can be represented by different number of cells. One with uniform cell diameter and the other with non-uniform cell diameter. It can be seen that both sets have the same shape and size, albeit the number of cells. This is the central idea we use in the proposed adaptive algorithm.

The proposed algorithm seeks to alleviate the cell growth draw back by subdividing only a subset of the cells obtained after the selection step. The question remains what criteria to use to select the cells to be subdivided. Indeed, an earlier work on set-oriented methods for analysis of autonomous dynamical system used approximations of the Sinai-Bowen-Ruelle (SBR) measures as the criterion to adaptively select the cells for subdivision \cite{DELLNITZ1998}. We however take a different approach since it could be equally challenging to compute the SBR measures. 

Let $\mathcal{B}(\mathcal{C}_{d_k})$, $\mathcal{N}(\mathcal{B}(\mathcal{C}_{d_k}))$ and $\mathcal I(\mathcal{C}_{d_k})$ be the boundary, neighborhood of the boundary and interior cells respectively, at iteration $k$ of the standard GIS algorithm. In line with the earlier explanation, the proposed algorithm selects the boundary cells $\mathcal{B}(\mathcal{C}_{d_k})$ for subdivision. However, since the boundary of the largest control invariant set $R_X$ or the region where it lies is not precisely known, the algorithm includes additional cells within the neighborhood of the boundary cells $\mathcal{N}(\mathcal{B}(\mathcal{C}_{d_k}))$. The selection of the boundary cells $\mathcal{N}(\mathcal{B}(\mathcal{C}_{d_k}))$ is based on the $N$-nearest neighbor ($N$-NN) supervised machine learning algorithm,  and is controlled by a parameter $N$. This will be described later in this section. The adaptive subdivision step at each iteration of the algorithm involves three main steps. The first step involves selecting the boundary cells $\mathcal{B}(\mathcal{C}_{d_k})$. The second step involves selecting a neighborhood of the boundary cells $\mathcal{N}(\mathcal{B}(\mathcal{C}_{d_k}))$. Finally, both group of cells are subdivided.

We begin this section by first describing the procedure for selecting the boundary. Thereafter, we described how the neighborhood of the boundary cells are selected.
\subsubsection{Boundary selection}

\begin{figure}[tbp] 
    \center{\includegraphics[width=0.4\columnwidth]{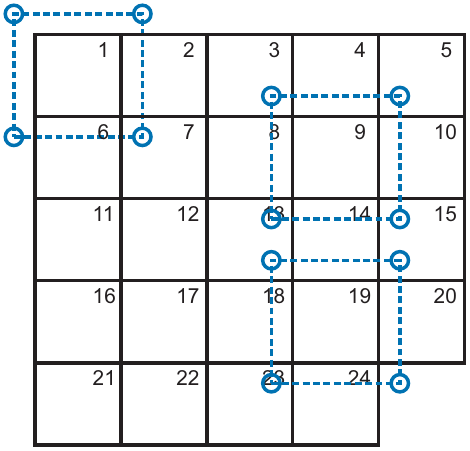}}
    \caption{\label{fig:boundary_selection} Graphical illustration of the process for selecting the boundary. The dashed blue lines represents the cell enlargement and the circles represent the vertices of the enlarged cells.}
\end{figure}

The goal of the boundary selection step is to find $\mathcal{B}(\mathcal{C}_{d_k})$.  Consider the cell $B_i \in \mathcal{C}_{d_k}$ at iteration $k$ of the algorithm. The Cell $B_i$ is first enlarged by a small factor $\delta$. Afterwards, the vertices of the enlarged cell are selected. The idea is that, each vertex of the enlarged cell must intersect a neighboring cell if it is an interior cell $\mathcal I(\mathcal{C}_{d_k})$, otherwise it is a boundary cell $\mathcal B(\mathcal{C}_{d_k})$. This is illustrated in Figure \ref{fig:boundary_selection}. 

In Figure \ref{fig:boundary_selection}, Cells $B_1$ to $B_{24}$ are the cells that constitute the cells to be subdivided $\mathcal{C}_{d_k}$. This implies that Cell $B_{25}$ has been removed from the previous iteration of the algorithm. For illustration purposes, the cell enlargement is shown for Cells $B_1$, $B_9$ and $B_{19}$ as the blue dashed rectangles. The vertices of the enlarged cells are selected after the enlargement. This is shown as the blue small circles. The expectation is that if a cell in an interior cell, then all four selected vertices must intersect a neighboring cell. For Cell $B_1$, it can be seen that only one vertex intersects a neighboring cell, that is Cell $B_7$. The other vertices do not intersect any neighboring cell. This makes Cell $B_1$ a boundary cell. Similarly for Cell $B_{19}$, three vertices of the enlarged cell intersect neighboring cells with one vertex not intersecting any cell. This implies that Cell $B_{19}$ is also a boundary cell. Finally, following the same procedure for Cell $B_9$, it can be seen that all the vertices intersect its neighboring cells. This makes it an interior cell. 
A summary of the boundary selection algorithm is summarized in Algorithm \ref{alg:boundary_selection}.

\begin{remark}
    It is possible that a cell which is supposed to be a boundary cell is not selected. This is an edge case. While edge cases in the boundary selection process are not expected, they can occur. For example, in Figure \ref{fig:boundary_selection}, if Cell 4 is not present, then Cell 9 is supposed to be a boundary cell. However, Cell 9 will not be selected as a boundary cell because all the vertices of the enlarged cell satisfy the criterion for it to be an interior cell. In this case, the procedure can be modified such that points along the edges of the enlarged cell are also included. This may however impact the computational speed. 

    Moreover, the selection of the neighborhood of the boundary will automatically resolve such edge cases. This will be described shortly in the next subsection.
\end{remark}

\begin{algorithm}

\caption{Selection of boundary cells} \label{alg:boundary_selection}

\KwIn{Cells to be subdivided $\mathcal{C}_{d_k}$, boundary cells $\mathcal{B}(\mathcal{C}_{d_k})$ }
\KwOut{Boundary cells $\mathcal B(\mathcal{C}_{d_k})$}
$\mathcal{B}(\mathcal{C}_{d_k}) \leftarrow \emptyset$ \tcp{Initialization}
\For{$ B_i \in \mathcal{C}_{d_k}$}{
    
    Enlarge the cell $B_i$ \\
    Select the vertices of the enlarged cell \\

    \If{ all the vertices do not intersect cells in $\mathcal{C}_{d_k}$}{ 
        Add $B_i$ to the collection $\mathcal{B}(\mathcal{C}_{d_k})$ \tcp{$B_i$ is a boundary cell}   
    }
}
\textbf{return $\mathcal B(\mathcal{C}_{d_k})$}

\end{algorithm}

\subsubsection{Selection of neighborhood of boundary cells}

As mentioned earlier, since the location of the boundary of the largest control invariant set $R_X$ is unknown in advance, a neighborhood of the boundary cells is also selected for subdivision. The implications of selecting or not selecting the neighborhood of the boundary cells will be demonstrated in the results section. Therefore, following the boundary cell selection, the neighborhood cells of each boundary cell are also selected for subdivision. This is achieved using the $N$-nearest neighbors ($N$-NN) of a point algorithm. $N$ is a parameter which determines the number of neighboring cells to be selected. This ultimately determines how far from the boundary cells we want to move into the interior of the set. 

\begin{figure}[tbp] 
    \center{\includegraphics[width=0.4\columnwidth]{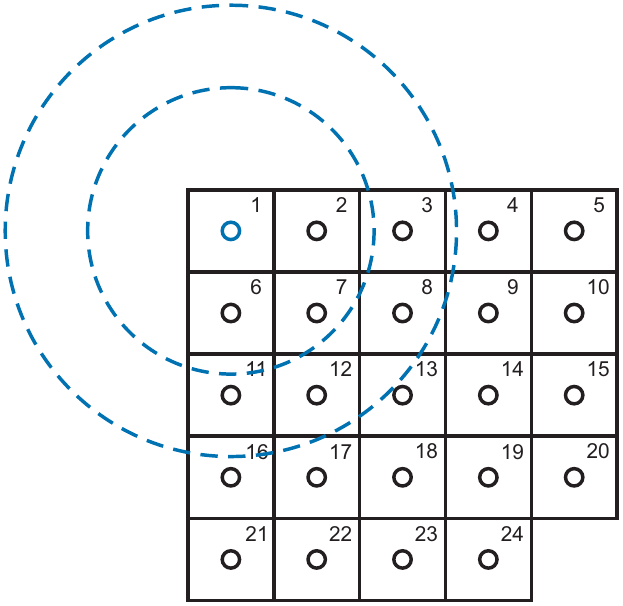}}
    \caption{\label{fig:n_nearest_neighbors} Graphical illustration of the process for selecting the neighborhood of the boundary cells. The center of the cells are indicated with the solid circles. The two dashed circles show the $N$-nearest neighbors of Cell $B_1$ for different $N$. The smaller dashed circle corresponds to an $N$ of 3 while the larger dashed circle corresponds to an $N$ of 7. Thus, for the smaller dashed circle, the three cells namely, $B_2$, $B_6$ and $B_7$ are selected. The Cells $B_2$, $B_6$, $B_7$, $B_3$, $B_8$, $B_{12}$ and $B_{11}$ are selected for the larger dashed circle.}
\end{figure}

$N$-NN is a supervised machine learning technique which is used to solve classification and regression problems. Given a point $p$, it selects the $N$ neighboring points of $p$ using the distances from the point. This is implemented in the adaptive GIS as follows. First, the center of the cells are selected. Then for each boundary cell, the $N$-nearest neighboring points are selected. Figure \ref{fig:n_nearest_neighbors} shows how the $N$-NN algorithm is used to select the $N$-neighborhood of the Cell $B_1$. 

\begin{remark}
  The presence of dataset imbalance and outliers can significantly affect the $N$-NN algorithm. However, in this work we do not expect these situations to happen since the outliers are absent and there is no dataset imbalance.
\end{remark}





    




\subsection{Efficient parallelization with GPU}

Because of the complicated nature of the GIS algorithm, parallelization with GPU is not trivial. For example, if the parallelization is not done properly, it can lead to computational inefficiencies due to excessive communications between the processors and GPU. Several issues need to be addressed for efficient parallelization:
\begin{itemize}
    \item Load-balancing: Parallelization cannot be achieved if the tasks to be completed are not distributed fairly among the compute cores. We use load-balancing to improve the parallelization in the graph creation step.
    \item Batching: The number of cells in the GIS algorithm can grow quickly. Loading and unloading data to and from the GPU can significantly degrade the benefits of using the GPU. We use batching to load a number of cells onto the GPU at time to maximize the use of the GPU.
\end{itemize}
Using appropriate batching and load-balancing schemes can significantly reduce the data traffic between the CPU cores and the GPU. In this section, we outline the details of the parallelization of the GIS algorithm.

\begin{figure}[tbp] 
    \center{\includegraphics[width=0.7\columnwidth]{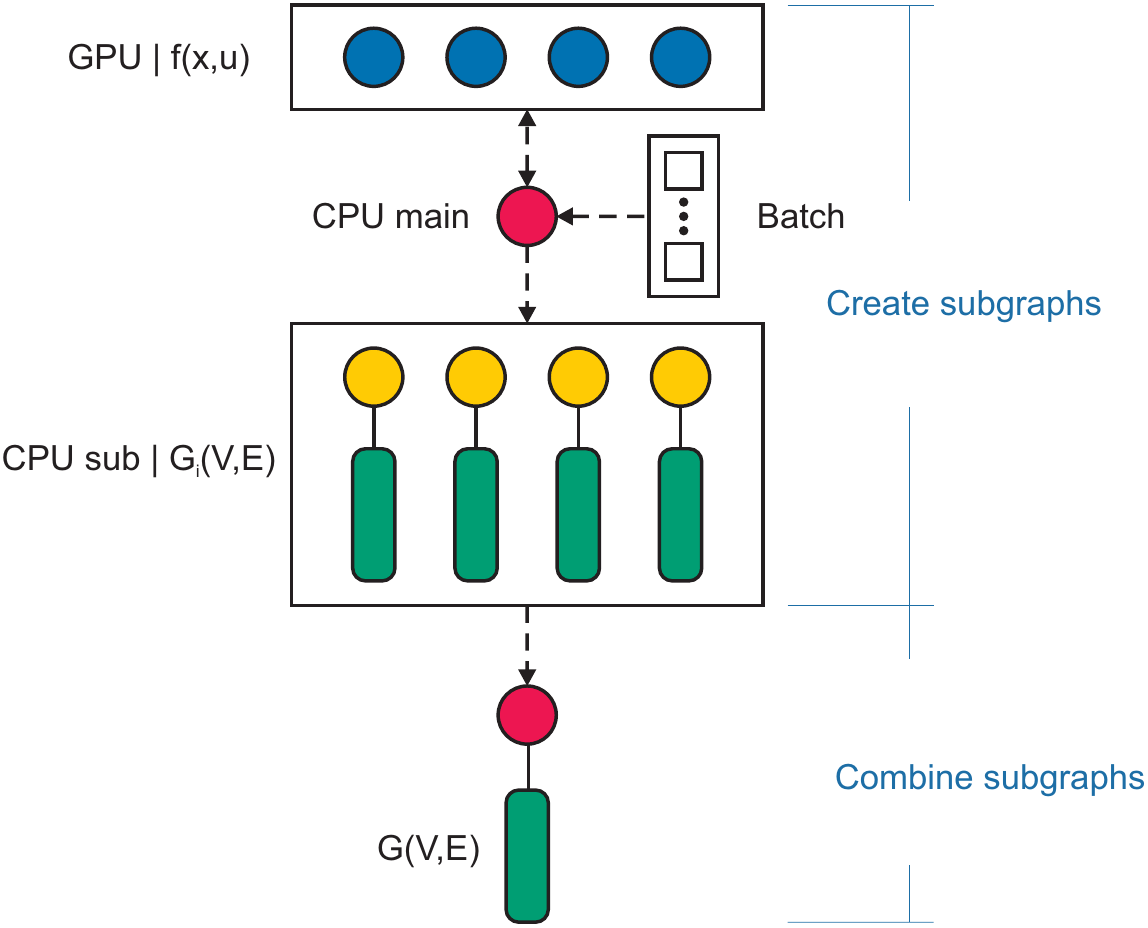}}
    \caption{Coordination of CPU and GPU, and data flow in the parallelized graph construction step of the improved GIS algorithm \label{fig:parallel_graph_construction} }
\end{figure}

In the GIS algorithm, one step forward mappings need to be computed for each cell. This can be done for each cell independently. Thus, the same instruction is used for each cell. This type of parallelization is known as data parallelization. GPUs are particularly suited for this kind of operation than CPUs. This is because GPUs have a highly parallel structure which make them more efficient for algorithms that process large chunks of data in parallel. A typical GPU usage sequence involves 
\begin{enumerate}
\item loading the data from the CPU to the GPU
\item performing the computation on the data
\item offloading the data from the GPU to the CPU
\end{enumerate}
The main speed up when using the GPU is from the second step. Steps (1) and (3) are the main bottlenecks when using a GPU. Thus, frequent loading and offloading of data to and from the GPU can significantly overshadow the gains made in Step (2). To address this problem, we load the cells in batches. This is in contrast to the sequential GIS algorithm where the one step forward mapping is computed one cell at a time. By loading a number of cells at a time (batch), the communication frequency between the CPU and GPU is reduced significantly. The number of cells to load from the CPU to the GPU depends on the available memory on the GPU.

Let $B_i^+$ be the cells that have an intersection with $F(B_i)$, that is, $B_i^+ =: \{ B_j | B_j \cap F(B_i) \neq \emptyset, B_j \in \mathcal{C}_{d_k} \}$. In the parallelized GIS algorithm, the main CPU passes a batch of $n$ cells $B_i \hdots B_{i+n}$ to the GPU which then computes and returns the images $F(B_i) \hdots F(B_{i+n})$.  Once the main CPU receives the data from the GPU, it creates and distributes the data across a number of subprocesses. Each subprocess finds the corresponding $B_i^+$ using the $F(B_i)$ information. To avoid race condition when each subprocess writes the edge data into the same graph, a subgraph  $G_i(V,E)$ is created for each subprocess $i$. This continues until all the cells in $\mathcal{C}_{d_k}$ are exhausted. Then in the second step, the graphs are merged into a single graph $G(V,E)$. This is then passed to the graph analysis step for processing. Figure \ref{fig:parallel_graph_construction} shows the data flow and how the main CPU coordinates with the GPU and subprocesses.

\subsection{Convergence issues}

In this section, we briefly discuss the implications of the modifications to the GIS algorithm on the convergence of the sets to the largest control invariant set $R_X$. We note that parallelization of the graph construction step does not affect convergence to $R_X$ in anyway. This is because other than speeding up the construction of the graph, the graph is not modified in any way. We therefore focus on the implication of the adaptive subdivision on the convergence to the algorithm to $R_X$. 

In the adaptive subdivision technique, only a subset of the cells are subdivided. This is in contrast to the standard GIS algorithm where all the cells are subdivided at each iteration. However, we note that this modification does not affect the convergence of the algorithm other than the speed of convergence. As an illustration, let us assume the worst case scenario where the boundary of the largest control invariant set lies somewhere deep in the interior of the state constraint. If only the boundary cells are selected and subdivided, then the algorithm will spend majority of the time refining cells which do not contain the boundary of $R_X$. The algorithm will keep refining the irrelevant cells until they are sufficiently small, only to remove those cells before moving to the next boundary cells which are much coarser since they have not been subdivided. This continues until the boundary cells which contain the boundary of $R_X$ are eventually located. While the cell growth is significantly reduces, the rate of convergence also reduces. This is certainly different from the standard GIS algorithm where all the cells are subdivided and therefore the boundary of $R_X$ can be found much faster. The addition of the neighborhood of the boundary cells helps to balance the trade off between faster convergence and cell growth rate. This will be demonstrated in the results section.

\section{Results} \label{sec:results}

In this section, we test the effectiveness of the modifications to the standard graph-based control invariant set computation algorithm. First, we consider the effects of the adaptive subdivision modification on the standard algorithm. Then, we consider the impact of the parallelization. Both tests were performed using a nonlinear continuously stirred tank reactor example. In all these cases, the computations were run on workstation with the following configuration: a quadcore Intel i7-4720HQ CPU with frequency of 2.6 GHz, 16 GB of random access memory (RAM), and Nvidia GeForce GTX 960M GPU with 2 GB video RAM.

\subsection{Process description}
We consider a well-mixed continuously stirred tank reactor (CSTR) where a first-order irreversible reaction of the form $A \rightarrow B$ takes place. Because the reaction is exothermic, thermal energy is removed from the reactor through a cooling jacket. Assuming constant volume reaction mixture, the following differential equations are obtained based on energy balance and mass balance for component $A$.
\begin{subequations} \label{eqn:cstr_equations}
    \begin{align} 
        \frac{dc_A}{dt} ={} & \frac{q}{V}(c_{Af} - C_A) - k_0 \exp(-\frac{E}{RT})c_A \\
        \frac{dT}{dt} ={} & \frac{q}{V}(T_{f} - T) +  \frac{-\Delta H}{\rho c_p} k_0 \exp(-\frac{E}{RT})c_A + \frac{UA}{V \rho c_p}(T_c - T)
    \end{align}
\end{subequations}
where $c_A$ and $T$ denote the reactant concentration and temperature of the reaction mixture in $mol/L$ and $K$ respectively, $T_c$ denotes the temperature of the coolant stream in $K$, $q$ denotes the volumetric flow rate of the inlet and outlet streams of the reactor in $L/min$, $c_{Af}$ denotes the concentration of reactant $A$ in the feed stream, $V$ denotes the volume of the reaction mixture, $k_0$ denotes the reaction rate pre-exponential factor, $E$ denotes the activation energy, $R$ is the universal gas constant, $\rho$ is the density of the reaction mixture, $T_f$ is the temperature of the feed stream, $c_p$ is the specific heat capacity of the reaction mixture, $\Delta H$ is the heat of reaction and $UA$ is the heat transfer coefficient between the cooling jacket and the reactor. The values of the parameters used in the simulations are listed in Table \ref{tb:parameters}.
\begin{table}[tbp]
  \begin{center}
  \caption{Table of parameter values}\label{tb:parameters}
    \begin{tabular}{ccl} \hline
    Parameter & Unit & Value \\\hline
    $q$ & $L/min$ & $100.0$ \\
    $V$ & $L$ & $100.0$ \\
    $c_{Af}$ & $mol/L$ & $1.0$ \\
    $T_f$ & $K$ & 350.0 \\
    $E/R$ & $K$ & $8750.0$ \\
    $k_0$ & $min^{-1}$ & $7.2 \times 10^{10}$ \\
    $-\Delta H$ & $J/mol$ & $5.0 \times 10^4$ \\
    $UA$ & $J/min\cdot K$ & $5.0 \times 10^4$ \\
    $c_p$ & $J/g\cdot K$ & $0.239$ \\
    $\rho$ & $g/L$ & $1000.0$ \\ \hline
    \end{tabular}
  \end{center}
\end{table}
The nonlinear model of (\ref{eqn:cstr_equations}) is discretized using a step-size  $h = 0.1$ $min$ to obtain a discrete-time nonlinear state space model of the following form in System \eqref{eqn:system} where $x=[C_A ~ T]^T$ is the state vector and $u=T_c$ is the input. The state and input are assumed to be subject to the following hard constraints: $0.0 \leq x_1 \leq 1.0$, $345.0 \leq x_2 \leq 355.0$, $285.0 \leq u \leq 315.0$.

\subsection{Adaptive subdivision results}

\begin{figure}[tbp] 
    \center{\includegraphics[width=0.6\columnwidth]{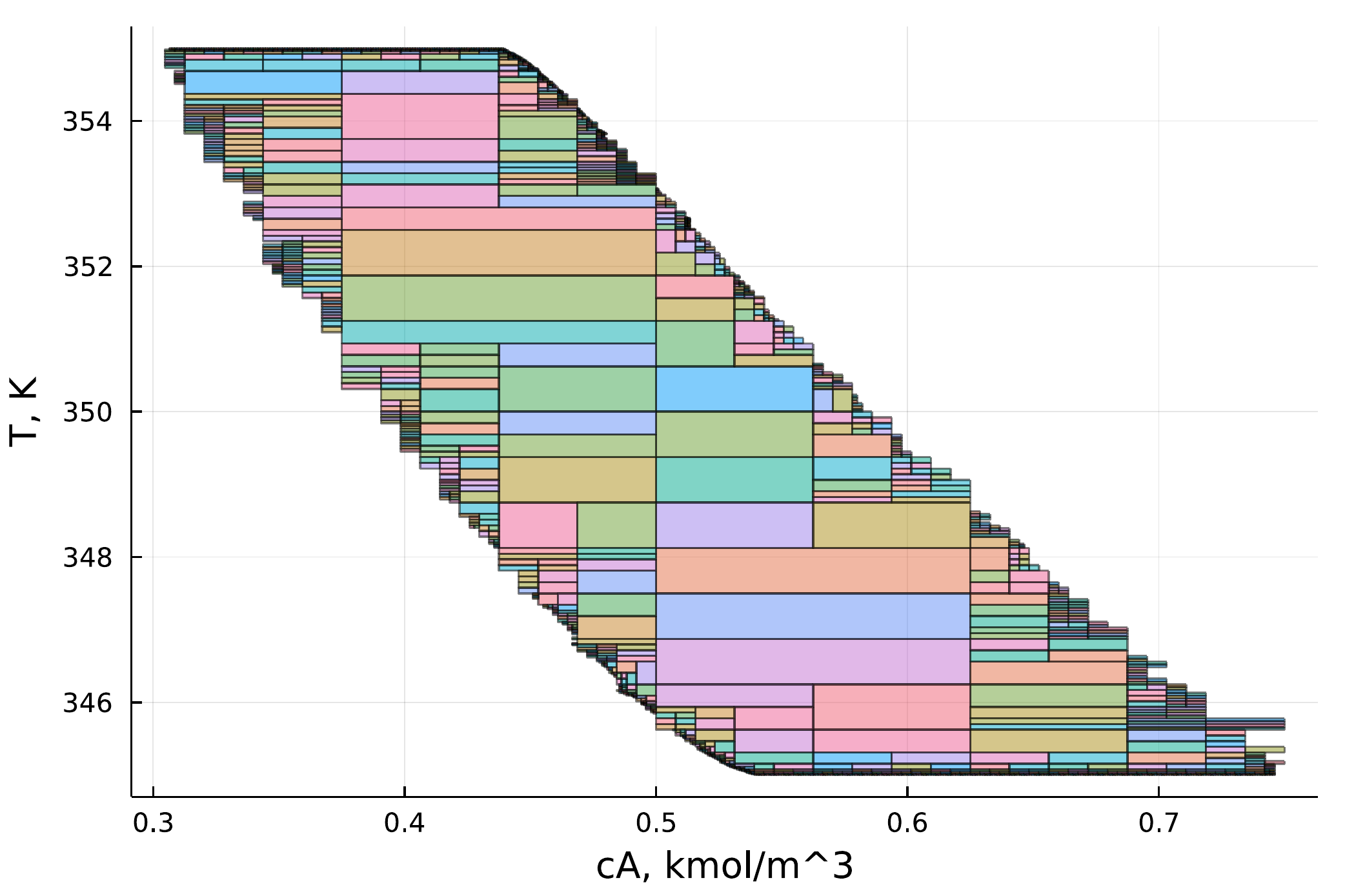}}
    \caption{\label{fig:adaptive_subdivision} Sample plot of the cells after 20 iterations of the adaptive algorithm with $N=0$ }
\end{figure}

In this section we present the results of the adaptive subdivision without the parallelization. Figure \ref{fig:adaptive_subdivision} shows a sample output of the algorithm after 20 iterations with $N=0$. This means that only the boundary cells are selected for subdivision. This value of $N$ was chosen to ensure that the number of cells is not too large to slow down the plotting of the figure. As expected, it can be seen that the cells are finer at the boundaries and coarser at the interior of the set.

\begin{figure}[tbp] 
    \center{\includegraphics[width=0.6\columnwidth]{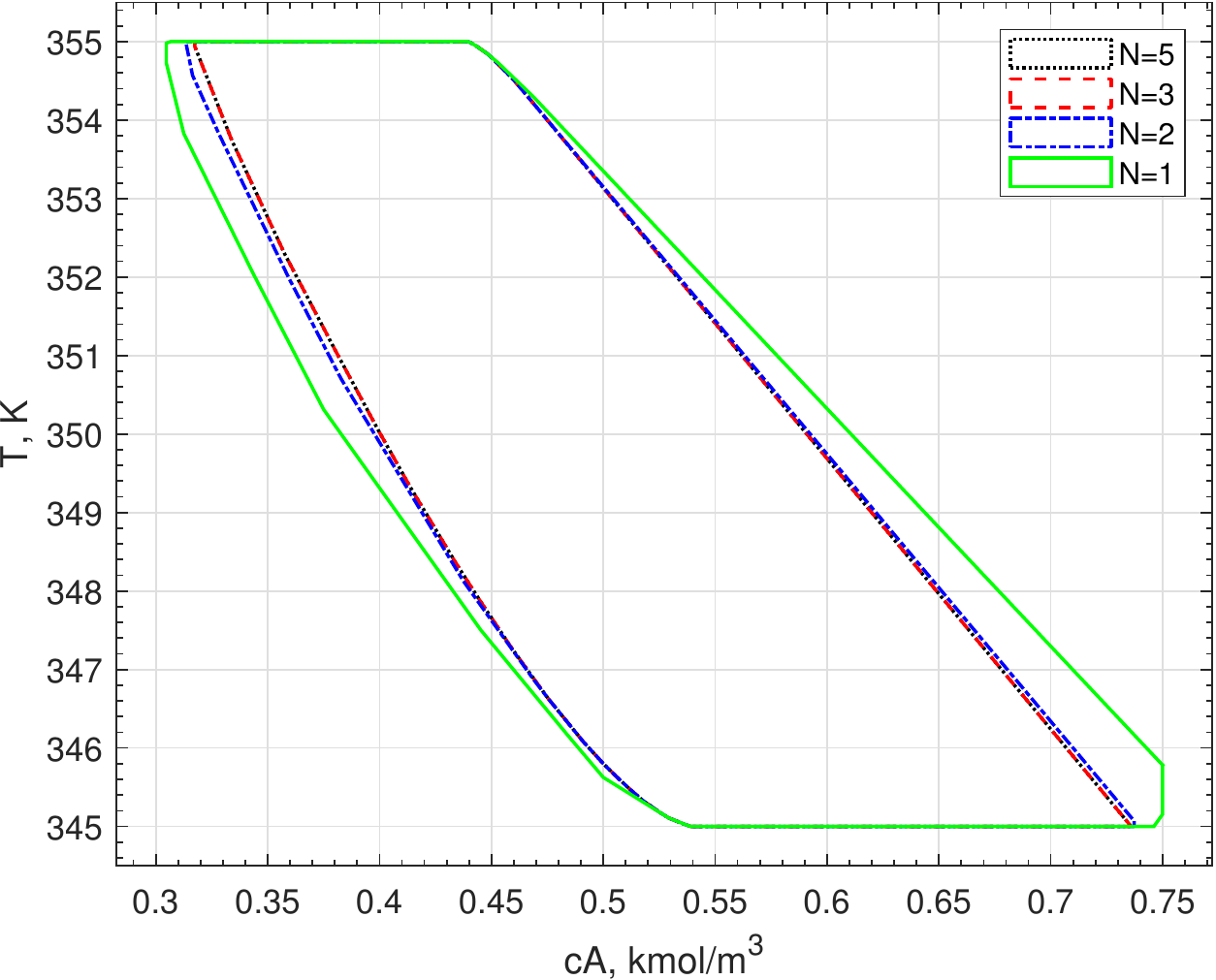}}
    \caption{\label{fig:adaptive_cell_division_cstr_set_comparison} Sets for different $N$ after 20 iterations. The invariant sets in the figure were obtained by finding the convex hull of the final cells after the algorithm  }
\end{figure}

\begin{figure}[tbp] 
    \center{\includegraphics[width=0.6\columnwidth]{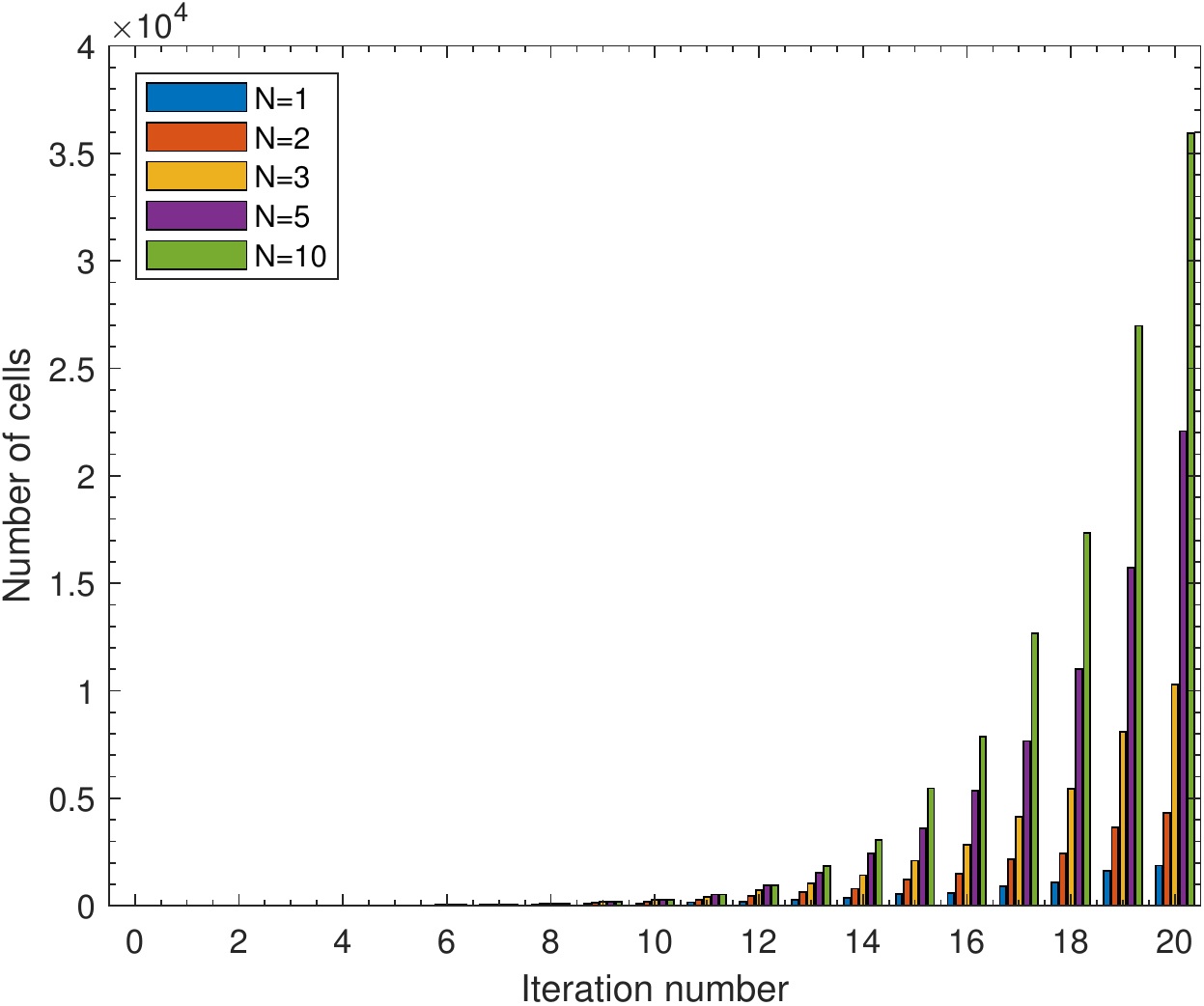}}
    \caption{\label{fig:adaptive_cell_division_cstr_cell_generation} Number of cells generated at each iteration of the algorithm with different $N$.}
\end{figure}

\begin{figure}[tbp] 
    \center{\includegraphics[width=0.6\columnwidth]{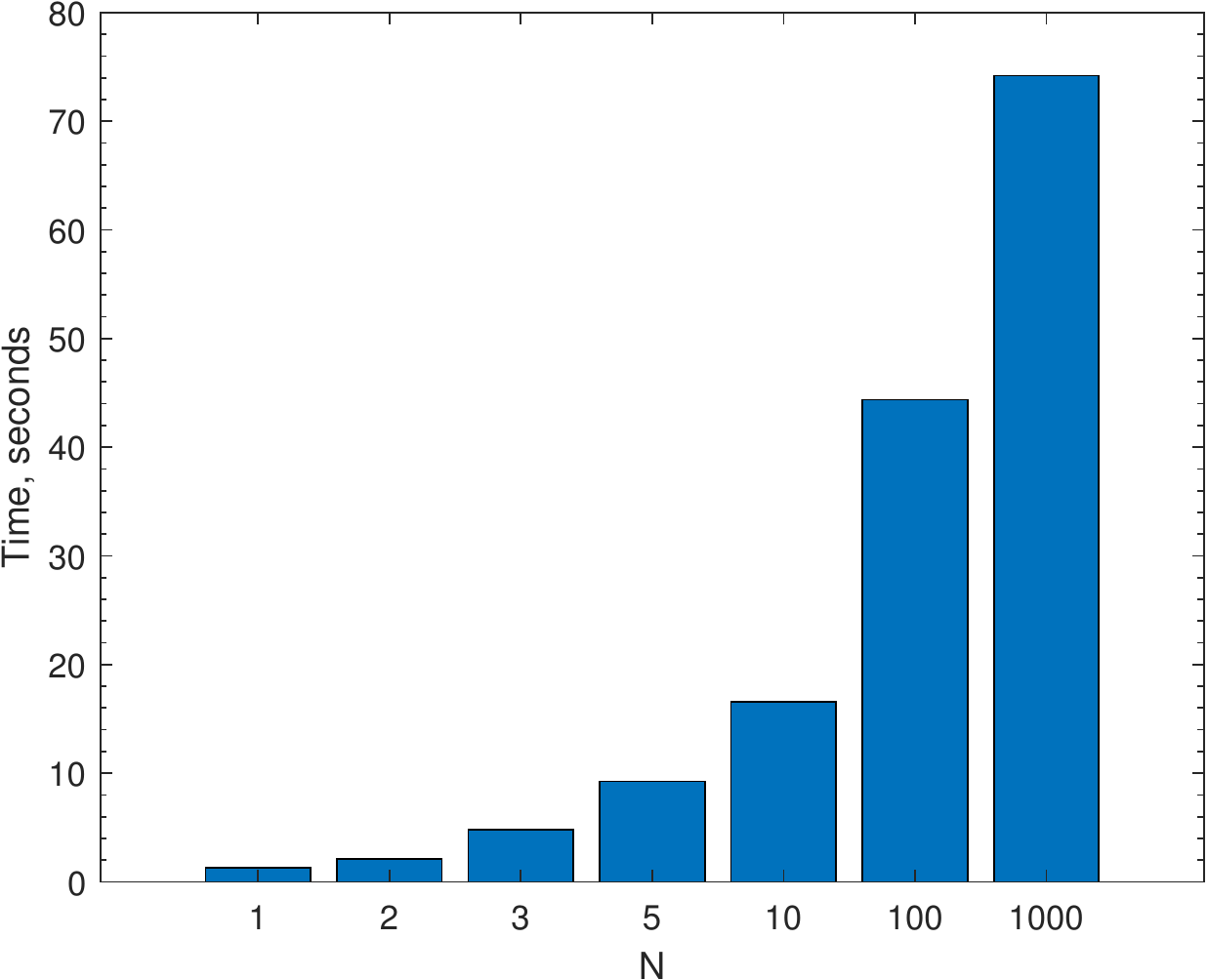}}
    \caption{\label{fig:adaptive_cell_division_cstr_computation_times} Computation times for different $N$. }
\end{figure}

In the next set of computations, we varied the parameter $N$ and recorded the number of cells generated at each iteration as well as the computation times. Figures \ref{fig:adaptive_cell_division_cstr_set_comparison}, \ref{fig:adaptive_cell_division_cstr_cell_generation} and \ref{fig:adaptive_cell_division_cstr_computation_times} show the convex hull of the sets generated, the number of cells generated and the computation times after 20 iterations respectively for different values of $N$. It can be seen in Figure \ref{fig:adaptive_cell_division_cstr_set_comparison} that the parameter $N$ affects the speed of convergence of the algorithm to the largest control invariant set $R_X$. 
For the same number of iterations, that is 20, the computation with a higher value of $N$ converges faster to $R_X$ compared that with a smaller value. This implies that a higher number of iterations is needed for the set to converge to $R_X$ when for example, $N=1$. As explained in Section \ref{sec:efficient_gis_algorithm}, when a small value of $N$ is used, the computation focuses on the boundary cells while searching for the boundary of $R_X$. Thus, more time is spent refining the cells in areas which do not contain the boundary of $R_X$. However, as the value of $N$ increases, more cells are selected. This means that larger areas within the domain of interest is explored in the search for the largest control invariant set. Hence, the selection of the parameter $N$ is not a trivial task. The choice of the value of $N$ depends on the properties of the system under study.

In this example, it can be seen from Figure \ref{fig:adaptive_cell_division_cstr_set_comparison} that the optimal value of $N$ is 3 since there's no difference between the set when $N=3$ and $N=5$. Furthermore, taking a closer look at the number of cells generated and the computation times (in Figures \ref{fig:adaptive_cell_division_cstr_cell_generation} and \ref{fig:adaptive_cell_division_cstr_computation_times} respectively), the number of cells is significantly reduced at $N=3$. Also, at $N=3$, the computational savings is 8 times that of $N=1000$ which represents the case when all the cells are subdivided at each iteration.

\subsection{Parallelization results}

\begin{figure}[tbp] 
    \center{\includegraphics[width=0.6\columnwidth]{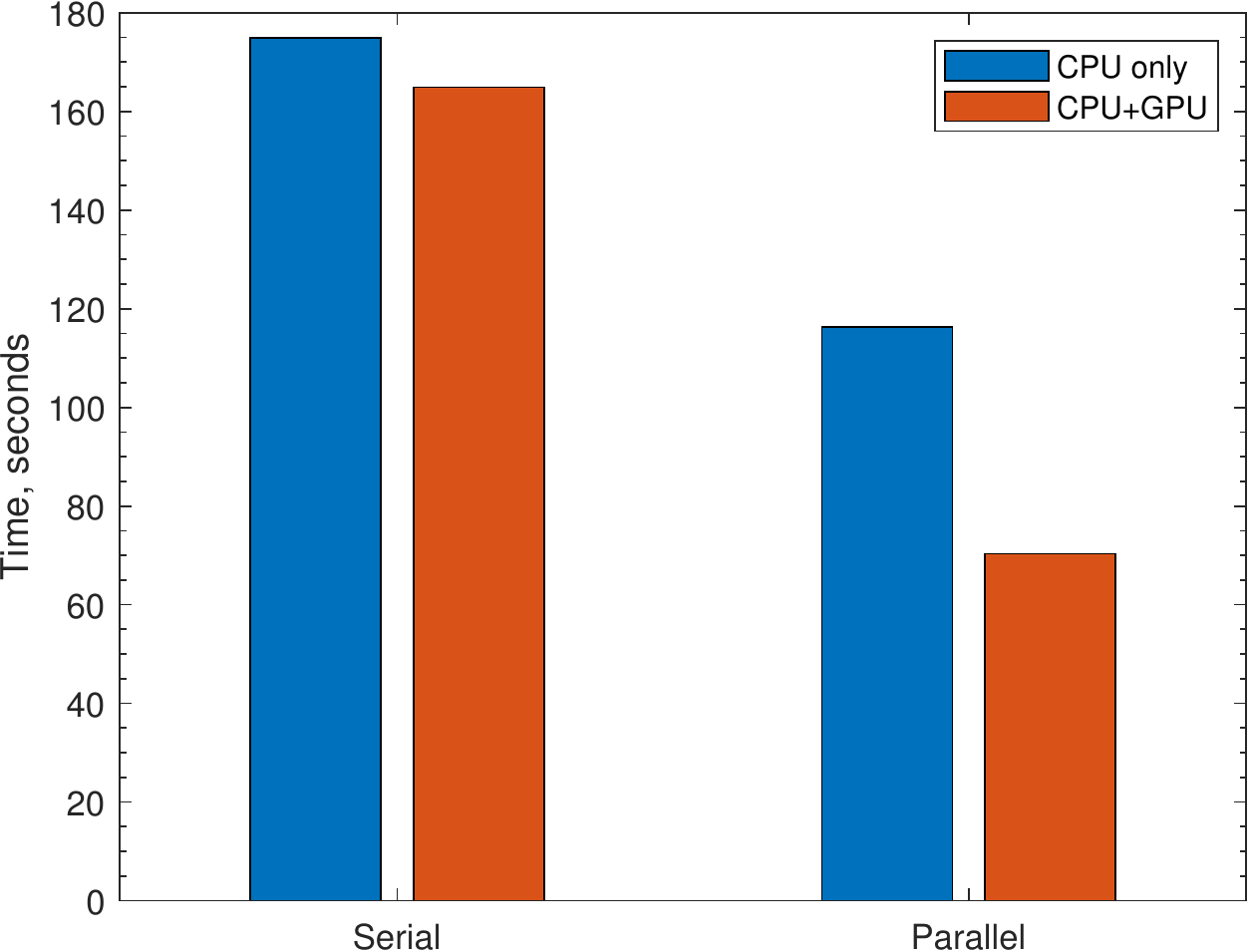}}
    \caption{\label{fig:cpu_gpu_cstr_computation_times} Comparison of computation speed for both serial and parallel computation, with and  without GPU usage.}
\end{figure}

In this section, we compare the computational speed improvements for the parallelization of the graph construction step of the algorithm. In this set of simulations, all the cells were subdivided without using the adaptive cell subdivision. This is to ensure that only the effect of the parallelization observed. We also considered the effect of using the GPU on the computation speed. Figure \ref{fig:cpu_gpu_cstr_computation_times} summarizes the results of the computation of the largest control invariant set using the parallelized algorithm. It can be seen that in both cases, the parallelization sped up the computation with a much improved speed, in the GPU case. The number of cells in a batch was selected as 1024 in the case where GPU is used.

\section{Concluding remarks} \label{sec:concluding_remarks}

In this work, we have presented an improved and efficient graph-based invariant set algorithm for computing approximations of the largest control invariant set of constrained controlled nonlinear systems. We first critically analyzed the computational complexity of the standard GIS algorithm. It was observed that the graph construction and the subdivision steps have significant impacts on the overall time complexity of the algorithm. Thus, we proposed two methods to improve the algorithm namely, adaptive subdivision and parallelization of the graph construction step. We demonstrated the efficacy of the improved algorithm using a nonlinear continuously stirred tank reactor. It was observed that the adaptive subdivision method only affects the speed of convergence to the largest control invariant set and not the convergence itself. Furthermore, the adaptive subdivision improved the speed of the algorithm by about 8x that of standard algorithm. Also, the parallelization of the graph construction step improved the computation speed by about 3x that of the standard algorithm.

In the future, we would look at further improving the computational speed by using a system decomposition-based approach. It will also be interesting to quantify the rate of convergence in the adaptive scheme. This can be used to tune the parameter $N$ since its selection is not trivial.

\section{Acknowledgement} \label{sec:acknowledgement}
This work is supported in part by the Natural Sciences and Engineering Research Council of Canada.


\end{document}